\documentstyle[fullpage,aps,epsf]{revtex}

\begin{document}
\title{Phonon Scattering by Breathers\\
in the Discrete Nonlinear Schr\"{o}dinger Equation}

\author{
Sungsik Lee\footnote{Electronic address: ssong@anyon.postech.ac.kr},
Julian J.-L. Ting\footnote{Electronic address: jlting@anyon.postech.ac.kr}
and Seunghwan Kim\footnote{Electronic address: swan@postech.ac.kr}}
\address{
Department of Physics,
Pohang University of Science and Technology,
San 31 Hyojadong, Pohang, Korea 790-784}
\date{Phys. Rev. E accepted June 24, 1998}

\maketitle

\begin{abstract}
Linear theory for phonon scattering by discrete breathers in the 
discrete nonlinear Schr\"{o}dinger equation 
using the transfer matrix approach is 
presented. 
Transmission and reflection coefficients are obtained as a function 
of the wave vector of 
the input phonon.
The occurrence of a nonzero transmission,
which in fact becomes perfect for a symmetric breather,
is shown to be connected with 
localized eigenmodes thresholds. 
In the weak-coupling limit, perfect reflection are shown to exist,
which requires two scattering channels.
A necessary 
condition for a system to have a perfect reflection is also considered in 
a general context.
\end{abstract}

\section{introduction}
Breathers, which also called oscillating solitons, 
are time periodic and spatially localized excitations
ubiquitously found on  translationally invariant coupled nonlinear 
lattice\cite{R1,R16,R17}. 
It was introduced first in the context
of the sine-Gordon equation in 1974\cite{AKNS}.
It requires practically no activation energy and thus bridge the gap
between the highly nonlinear modes and the linear phonon modes.
Such long-lived spatially localized excitation has fundamental 
significance in the dynamics of spatially extended systems.
It has been found that the discreteness of a system plays
an essential role on the stability of breathers in several systems\cite{MP}.
It introduces a cut-off frequency to the system so that the resonance between
the phonon band and the frequency of the discrete breather can be avoided.

However, breathers grow and decay due to their interaction with the
environment.
Some of such interactions which has been studied include the 
electron-breather interaction\cite{FK} and the interaction of breathers
with extended defect\cite{TP}.
The discrete breather can play a role of the scattering center 
affecting energy transport by scattering, 
absorbing  or radiating phonons.
Since the phonon is a major form of radiations for small amplitude excitations
in the nonlinear lattice,
it is important to characterize such breather-phonon interaction\cite{R3}.

Recently it has been found that
scattering properties of the breather are 
related to the structure 
of the internal modes of the breather itself\cite{R4}.
They studied the phonon scattering by a kink in the localized equilibria 
of the nearest neighbor interacting 
chain and showed that the scattering properties undergo
a drastic change, involving a perfect transmission,
as the localized eigenmode threshold is passed.
It was conjectured that this result on the static localized structure
could be extended to a wide class of 
systems with time-dependent localized structures such as breathers.
It has also been confirmed in the case of time periodic discrete breather in 
the Hamiltonian nearest neighbor 
chain both numerically and using the extended Levinson theorem\cite{R3}.

The purpose of this paper is to present the analysis of phonon scattering by 
breathers in the 
discrete nonlinear Schr\"{o}dinger 
(DNLS) equation using the transfer matrix method. 
We proved analytically that at the localized eigenmode 
threshold the transmission 
at the phonon band edge, such that the wave vector $k=0$,
is nonzero, which in fact becomes perfect in the case of symmetric breather.
This result is also confirmed by numerical calculations with transfer matrices.
The crucial difference of the DNLS equation from the localized
equilibria of the nearest neighbour chain is that it
allows two scattering channels\cite{R4}.
This allows occurrence of the perfect reflection whose 
existence is shown in detail in the weak-coupling limit.
A necessary condition for a system to have a perfect reflection is 
obtained in comparison with the case of the local equilibria in the 
classical Hamiltonian systems.
We end with discussions on extensions of our results to more general systems
including the time-periodic discrete breathers in the Hamiltonian chain.

\section{The Transfer Matrix Formulation}
There are many ways to discretize the
nonlinear Schr\"{o}dinger equation.
One of the widely studied version, called the DNLS equation, reads,
\begin{equation}
\label{eq1}
\dot\psi_{n} = i\{|\psi_{n}|^{2}\psi_{n} + \epsilon(\psi_{n-1}+\psi_{n+1})\},
\end{equation}
in which $\psi_{n}$ is the amplitude at site $n$ and $\epsilon$ is 
the coupling strength.
If one consider a solution of the form $\psi_{n}=a_{n}e^{i\omega t}$ 
for Eq.(\ref{eq1}), $a_{n}$'s satisfy
\begin{equation}
\label{eq2}
\omega a_{n} = a_{n}^{3} + \epsilon( a_{n-1} + a_{n+1} ).
\end{equation}
Without loss of generality, we could
choose $\omega>0$. 
We further consider only the 1-site breather, i.e. 
only one site is excited on the lattice.
Hence for zero-coupling 
the breather solution centered at the origin is given by 
$a_{0}=\sqrt{\omega}$ and $a_{n}=0$ for $n \neq 0$.

In the weak-coupling limit, i.e. if 
$\epsilon / \omega$ is small, 
we obtain from Eq.(\ref{eq2}) 
$a_{0}=\sqrt{\omega}( 1 - (\epsilon / \omega )^{2} + \cdots )$, 
          $a_{\pm 1}=\sqrt{\omega}
( \epsilon / \omega + \cdots )$, $\cdots$ etc.
This provides an exponentially localized amplitude profile of the discrete
breather
which can be continued from a zero-coupling solution\cite{R40}.
The reflection symmetry of the breather with respect to its center will be
shown to be important
in connection with perfect transmission.

The linearization      of Eq.(\ref{eq1}) around the breather
solution with $\phi_{n} = \psi_{n} - a_{n}e^{i\omega t} $ yields,
\begin{equation}
\label{eq4}
\dot\phi_{n} = i\{2a_{n}^{2}\phi_{n} + a_{n}^{2}e^{i2\omega t}\phi_{n}^{*} + \epsilon(\phi_{n+1}+\phi_{n-1})\}.
\end{equation}
Because of the complex conjugate operation on $\phi_{n}$, 
at least two frequency components are necessary 
for the solution of Eq.(\ref{eq4}).
Therefore, the DNLS equation can allow two scattering channels different from
the case of localized equilibria with a single scattering channel\cite{R4}.
If we assume 
$\phi_{n} = b_{n}e^{i(\omega+\Omega)t} + c_{n}e^{i(\omega-\Omega)t}$, 
with $b_n$ and $c_n$ complex, we obtained the following equations:
\begin{equation}
\label{eq5}
\left(\begin{array}{cc} \omega+\Omega-2 a_n^2 & 0\\ 
                          0  & \omega-\Omega-2 a_n^2 \end{array}\right)
\left(\begin{array}{c} b_{n} \\ c_{n} \end{array}\right)
-
\left(\begin{array}{cc} 0 & a_n^2 \\ 
                          a_n^2  & 0 \end{array}\right)
\left(\begin{array}{c} b_{n}^{*} \\ c_{n}^{*} \end{array}\right)
=
\epsilon
\left(\begin{array}{c} b_{n-1}+b_{n+1} \\ c_{n-1}+c_{n+1} \end{array}\right).
\end{equation}
With $ Y_{n}^{T} = (b_{n}, c_{n}, b_{n}^{*}, c_{n}^{*}) $, the above equation
can be written in a recursive form as 
\begin{equation}
\label{eq8}
\left( \begin{array}{l} Y_{n} \\ Y_{n-1} \end{array} \right)
=
M_{n}
\left( \begin{array}{cc} Y_{n+1} \\ Y_{n} \end{array} \right),
\end{equation}
in which $M_n$ is a function of $a_n$, $\epsilon$, $\omega$ and $\Omega$.
The matrix $M_n$ is called a transfer matrix, 
which maps a pair of linearized coordinates
on one site to another site\cite{TM}.
Applying Eq.(\ref{eq8}) successively from the 
site $N$ to the site $-N$, one has  
\begin{equation}
\label{eq10}
\left( \begin{array}{l} Y_{-N} \\ Y_{-N-1} \end{array} \right)
=
M
\left( \begin{array}{l} Y_{N+1} \\ Y_{N} \end{array} \right),
\end{equation}
with $M = M_{-N} \cdot M_{-N+1} \cdots M_{N-1} \cdot M_{N}$.

Far from the center of the breather, $a_{n}$ decays to zero, so that
Eq.(\ref{eq5}) 
becomes decoupled, with solutions of the form
$b_{n}=e^{ik_{b}n}$ and $c_{n}=e^{ik_{c}n}$, in which
$\cos{k_{b}} = (\omega+\Omega)/{2\epsilon}$ and 
$\cos{k_{c}} = (\omega-\Omega)/{2\epsilon}$.
The solutions of $b_n$ and $c_n$ define two possible scattering channels.
Each channel can be either a traveling (phonon) mode with a real wave vector
$k$
or an exponentially growing (decaying) mode with an imaginary $k$
depending upon the values of 
$|\omega \pm \Omega|/2\epsilon$.
Since $\epsilon$ and $\Omega$ occurs only in the form of ratios
$\epsilon/\omega$ and
$\Omega/\omega$, we can set $\omega =1$ and regard $\epsilon$ and $\Omega$ as 
dimensionless variables.
The boundaries of the phonon bands are given by $\omega = \Omega \pm 2 \epsilon$
and $-\Omega \pm 2 \epsilon$. These phonon bands divided the parameter space 
of $\epsilon$ and $\Omega$ into three regions 
according to the number of traveling channels 
as shown in Fig.\ref{fig1}.

\section{The Scattering Problem}
In a Levison theorem, the bound state property of a system is related to 
its scattering property\cite{LEV}.
In what follows, 
the property of scattering in region-$1$ was related 
to the property of localized mode in region-$0$.

Firstly, we consider the scattering problem in region-$1$.
In the standard scattering setup, 
a phonon is incident with normalized amplitude 
from the left side of the breather located at the origin.
We assume $\Omega>0$. The case with $\Omega<0$ can be handled similarly.

The question asked in a scattering problem is its transmission and reflection
coefficients.
It can be shown from the DNLS equation that
\begin{equation}
\label{conservation}
Im<\psi_{n+1}^{*}\psi_{n}>  =  Im<\psi_{n}^{*}\psi_{n-1}>,
\end{equation}
in which $<A>$ is the time average of $A$.
This flux-conservation relation makes 
the one-channel scattering process elastic, i.e.,
\begin{equation}
\label{eq13_8}
|t|^{2} + |r|^{2} = 1,
\end{equation}
in which $t$ and $r$ are transmission and reflection coefficients respectively. 

Suppose the channel $c_{n}$ is a traveling mode
with $k_c$ real, while $k_b = i \kappa_b (\kappa_b > 0)$ is imaginary.
The asymptotic solutions are of the form 
\begin{equation}
\label{eq13}
\begin{array}{ll}
         \begin{array}{lll}
            b_{n} &=& b^{-}_{l}e^{\kappa_{b}n} \\
            c_{n} &=& e^{ik_{c}n} + c^{-}_{l}e^{-ik_{c}n}  
         \end{array}   &
         \mbox{if $n \rightarrow - \infty$,} \\

         \begin{array}{lll}
            b_{n} &=& b^{+}_{r}e^{-\kappa_{b}n} \\ 
            c_{n} &=& c^{+}_{r}e^{ik_{c}n}  
         \end{array}   &
         \mbox{if $n \rightarrow \infty$,}
\end{array}
\end{equation}
in which the amplitude of the incident phonon on the
channel $c_n$ is set to unity.
The amplitudes for the transmitted and reflected phonons, 
$c_{r}^{+}$ and $c_{l}^{-}$, fully characterize the one-channel phonon 
scattering by a symmetric breather.
Since Eq.(\ref{eq5}) is not covariant under multiplication of  a
arbitrary phase factor $e^{i\theta}$ on $b_{n}$ and $c_{n}$, 
it may appear that the scattering properties 
depend on the phase of the input phonon. 
However, if $b_{n}$ and $c_{n}$ are solutions of Eq.(\ref{eq5}), 
so are $ib_{n}$ and $-ic_{n}$.
Therefore, transmission and reflection are independent of the phase of the 
input phonon in the one-channel scattering case.

Inserting Eq.(\ref{eq13}) into Eq.(\ref{eq8}), we obtain $8$ linear equations 
with $8$ variables to solve for, i.e. 
\begin{equation}
\label{eq13_5}
\left(   \begin{array}{c}
         b_{l}^{-}e^{-\kappa_{b}N} \\
         e^{-ik_{c}N}+c_{l}^{-}e^{ik_{c}N} \\
         b_{l}^{-*}e^{-\kappa_{b}N} \\
         e^{ik_{c}N}+c_{l}^{-*}e^{-ik_{c}N} \\
         b_{l}^{-}e^{-\kappa_{b}(N+1)} \\
         e^{-ik_{c}(N+1)}+c_{l}^{-}e^{ik_{c}(N+1)} \\
         b_{l}^{-*}e^{-\kappa_{b}(N+1)} \\
         e^{ik_{c}(N+1)}+c_{l}^{-*}e^{-ik_{c}(N+1)} 
         \end{array}
\right)
= M
\left(   \begin{array}{c}
         b_{r}^{+}e^{-\kappa_{b}(N+1)} \\
         c_{r}^{+}e^{ik_{c}(N+1)} \\
         b_{r}^{+*}e^{-\kappa_{b}(N+1)} \\
         c_{r}^{+*}e^{-ik_{c}(N+1)} \\
         b_{r}^{+}e^{-\kappa_{b}N} \\
         c_{r}^{+}e^{ik_{c}N} \\
         b_{r}^{+*}e^{-\kappa_{b}N} \\
         c_{r}^{+*}e^{-ik_{c}N} 
         \end{array}
\right).
\end{equation}
For $k_c = 0$, $1+c_l^-$ can be treated as one complex
variable. The resulting
$8$ homogeneous equations have a non-trivial solution only
on their null space. 

On the other hand, if both channels are non-traveling modes
with both $k_b = i\kappa_b (\kappa_b > 0)$ and 
$k_c = i \kappa_c (\kappa_c > 0)$
imaginary.
After discarding all exponentially growing parts, the asymptotic solutions
read,
\begin{equation}
\begin{array}{ll}
         \begin{array}{lll}
            b_{n} &=& b^{-}_{l}e^{\kappa_{b}n} \\
            c_{n} &=& c^{-}_{l}e^{\kappa_{c}n}  
         \end{array}   &
         \mbox{if $n \rightarrow - \infty$,} \\

         \begin{array}{lll}
            b_{n} &=& b^{+}_{r}e^{-\kappa_{b}n} \\ 
            c_{n} &=& c^{+}_{r}e^{-\kappa_{c}n}  
         \end{array}   &
         \mbox{if $n \rightarrow \infty$.}
\end{array}
\end{equation}
The localized eigenmode is given by a solution of 
\begin{equation}
\left(   \begin{array}{c}
         b_{l}^{-}e^{-\kappa_{b}N} \\
         c_{l}^{-}e^{-\kappa_{c}N} \\
         b_{l}^{-*}e^{-\kappa_{b}N} \\
         c_{l}^{-*}e^{-\kappa_{c}N} \\
         b_{l}^{-}e^{-\kappa_{b}(N+1)} \\
         c_{l}^{-}e^{-\kappa_{c}(N+1)} \\
         b_{l}^{-*}e^{-\kappa_{b}(N+1)} \\
         c_{l}^{-*}e^{-\kappa_{c}(N+1)} 
         \end{array}
\right)
= M
\left(   \begin{array}{c}
         b_{r}^{+}e^{-\kappa_{b}(N+1)} \\
         c_{r}^{+}e^{-\kappa_{c}(N+1)} \\
         b_{r}^{+*}e^{-\kappa_{b}(N+1)} \\
         c_{r}^{+*}e^{-\kappa_{c}(N+1)} \\
         b_{r}^{+}e^{-\kappa_{b}N} \\
         c_{r}^{+}e^{-\kappa_{c}N} \\
         b_{r}^{+*}e^{-\kappa_{b}N} \\
         c_{r}^{+*}e^{-\kappa_{c}N} 
         \end{array}
\right).
\label{eq13_6}
\end{equation}
Non-trivial solutions for this set of equations exist only on its null space.
At $k_c = 0$ the null spaces of Eq.(\ref{eq13_5}) and Eq.(\ref{eq13_6})
coincide. Therefore, at the band edge of the channel $c_n$, 
$k_c = 0$, where a localized
mode start to appear, we obtain non-zero phonon transmission, $c_r^{+}$.

We can further demonstrate that for symmetric breathers non-zero transmission
at $k_{c} =0$ at the threshold for localized mode generation becomes
perfect one:
Consider a phonon of unity amplitude incident from the left side of the 
breather at a small positive wave vector $k_{c}$. 
Near $k_{c}= 0$, 
this scattering configuration can  be extended continuously to a small 
negative value of $k_{c}$ 
since the governing  equations 
are smooth functions of $k_{c}$. 
Negating $k_{c}$ changes the direction of traveling phonons.
The configuration becomes an outgoing phonon of unity amplitude and 
an incoming phonon of amplitude $r (-k_{c} )$ on the left side of the breather,
and an incoming phonon with amplitude $t(-k_{c})$ on
the right side as shown in Fig.\ref{fig2}b.
For a symmetric breather, these incoming phonons with 
a wave vector $-k_c$ from the right side
of the breather generate
outgoing phonons of amplitude 
$r^{2}(-k_c)+t^{2}(-k_c)$ and $2 r(-k_c) t(-k_c) $ 
from the left and right sides of the breather,
respectively.
The situation in Fig.\ref{fig2}b and Fig.\ref{fig2}c must coincide for
$k_c = 0$.
Therefore, we obtain $r^{2}(0)+t^{2}(0)=1$ and $2 r(0) t(0) =0$, 
and conclude that there must be either perfect transmission or 
perfect reflection at $k_{c}=0$. 
This is a special property of the symmetric breather
since we assumed that the transmission and reflection coefficients for a 
phonon coming from the left side of the breather are the same as 
those from the right side of the breather. 

In general,
a scaterring process can be formulated 
as a set of homogeneous linear equations 
for the Fourier components of the state of the system 
at each site and casted into 
a transfer matrix form if the couplings between sites are local and linear.
Let $\vec Y_{n}$ be the state vector 
at site $n$ and $M_{n}$ be the transfer matrix
connecting $(\vec Y_{n+1}, \vec Y_{n})^T$ and 
$(\vec Y_{n}, \vec Y_{n-1})^T$ as in Eq.(\ref{eq8}).
If the dimension of $\vec Y_{n}$ is one, 
the whole vector $\vec Y_{n}$ represents the traveling phonon in 
the scattering process. 
If the incoming phonons from the left side of the breather 
are perfectly reflected,
the $\vec Y_{N}$ should be identical to zero as $N \rightarrow \infty$.
Because of the homogeneity of Eq.(\ref{eq10}), 
$\vec Y_{-N}$ should also be zero, 
which implies no non-trivial configuration of perfect reflection is possible.
If the dimension of $\vec Y_{n}$ is greater than one and some components 
represent the exponentially decaying parts, 
$\vec Y_{N}$ need not to be  zero
in the $N \rightarrow \infty$ limit in spite of the perfect reflection. 
It becomes possible to match non-zero $\vec Y_{-N}$, 
which represents the input and reflection of phonons, 
with $\vec Y_{N}$ by the transfer matrix. 
Therefore, the existence of additional channels 
with exponentially decaying 
parts is essential for 
a perfect reflection to occur in a linear scattering problem.
Hence, there is no perfect reflection in two-channel phonon 
scattering of the DNLS equation, i.e. in region-$2$ of Fig.\ref{fig1},
since all components of $\vec Y_{n}$ represent traveling phonons. 
This provides a  necessary condition for a linear-system to allow
a perfect reflection.

Our derivation does not depend on the specific form of the transfer 
matrix $M$. 
The result is entirely based on the fact that asymptotic solutions 
which are composed of traveling phonon parts and exponentially growing or 
decaying parts far from the center of the breather can be connected to each other 
by a transfer matrix. 
Therefore, the same procedure can be applied to the 
breather-phonon scattering problem in the nearest neighbor Hamiltonian chain 
to obtain a similar proof\cite{R3}.
A major difference is that the number of channels
in the latter case should be extended 
to infinity for each frequency component 
$\Omega+\omega_{b}$, in which $\Omega$ is the phonon frequency and 
$\omega_{b}$ is the breather frequency, 
in the case of Hamiltonian breather.

\section{Perturbative and Numerical Calculations}
The occurrence of the perfect reflection in the weak-coupling limit can be 
shown by a perturbation method. 
Retaining only terms of order $\epsilon / \omega $, 
we obtain the transfer matrix 
$M \approx M_{-1} \cdot M_{0} \cdot M_{1}$;
other $M_n$'s are discarded because they affect only the phase factors
in the order $ \epsilon / \omega $.
The transmission coefficient thus obtained is
\begin{equation}
\label{eq18}
|t|^{2} = 2 (\epsilon / \omega )^{4} \sin^{2}{2k_{c}}.
\end{equation}
The breather becomes opaque to the fourth power of the coupling strength.
A perfect reflection occurs at $k_{c}=\pi/2$ in the weak-coupling limit.
Numerical calculation,
which takes into account of several higher order terms 
of $\epsilon / \omega $, agrees with the perturbative result as
shown in Fig.\ref{fig3}.
Perfect phonon reflection by breathers in the DNLS equation is a 
property which persists for weak-coupling,
in contrast to what happens 
in the Hamiltonian nearest neighbor chain where the perfect 
reflection exist only for a positive interval of coupling strength 
$\epsilon$ for a given $\omega_{b}$. 

In the case of phonon scattering by a local equilibria kink in the 
nearest neighbor Hamiltonian chain, there is no perfect reflection\cite{R4}.
However, there occurs a perfect reflection in the Hamiltonian 
discrete breather\cite{R3}.
The critical difference between the case of a local equilibria kink and 
the case of Hamiltonian or the DNLS breathers is the 
minimum number of frequencies needed
for describing linearized motion at each site. 
In the case of the local equilibria, 
one Fourier component measuring the deviation from the equilibrium 
at each site is sufficient 
to describe the motion of each site 
because 
the kink in the 
local equilibria is a time independent defect. 
On the other hand,
the discrete breather in the Hamiltonian system has an intrinsic time scale 
$T_{b}$ which is the period of the breather. 
The intrinsic frequency $\omega_{b}=2\pi/T_b$ generates an infinite number 
of frequencies $\Omega+m\omega_{b}$
($m$ is an integer ) mixing with the phonon frequency $\Omega$ by nonlinear 
terms in the equation. 
For the DNLS case, only one more frequency is generated due to the 
complex conjugate operation in Eq.(\ref{eq4}).

With the transfer matrix method, 
the transmission and reflection coefficients can be calculated numerically 
in the same way as in the case of the local equilibria\cite{R4}.
The transfer matrix, $M$, is obtained numerically and 
the scattering coefficients are solved from Eq.(\ref{eq13_5}).
The number of sites in the scattering setup, $N$, 
in Eq.(\ref{eq13_5}) is chosen to allow a decay of the amplitude of the 
breather within the error bound, 
$10^{-2}$, in our calculation.
For $0.01 < \epsilon < 0.4$, $N$ needs not to exceed $5$ to maintain 
this bound.

The phase diagram for the localized eigenmodes and the transmission for 
various values of $\epsilon$ are obtained numerically in
Fig.\ref{fig4} and Fig.\ref{fig5}.
For small coupling, there is no perfect transmission.
A perfect transmission is generated from $k_{c}=0$ at $\epsilon=0.262$ 
where a symmetric localized eigenmode\cite{R35_1} is 
generated from the lower edge of the 
phonon band.
The value of $k_{c}$ where the perfect transmission occurs increases 
from zero until it reaches a maximum value.
Then the maximal wave vector
decreases and becomes zero again where another localized eigenmode, 
which is anti-symmetric at this time\cite{R35_1}, is generated.
These numerical results are consistent with predictions of the 
linear scattering analysis.

The return of $k_{c}$,
where the perfect transmission occurs, to 
zero after its initial creation and deviation from zero is 
due to the multi-site spreading of the breather envelope.
For a one-site breather, if only
$M_0$ is retained in calculating $M$, the location of the 
perfect transmission does not return to $k_{c}=0$ again;
which is consistent with the fact that
there is no generation of the anti-symmetric localized eigenmode.
In other words, the generation of a perfect transmission and 
a symmetric localized eigenmode 
is due to the single-site defect property of 
the discrete breather,
whereas the annihilation of the perfect transmission and the generation of 
an anti-symmetric localized eigenmode 
is due to the multi-site structure of the breather. 

The perfect reflection at nonzero $k_{c}$ is another special feature 
of the one-channel phonon scattering case.
It is important because the breather can be regarded as a 
perfect reflecting barrier for the phonon, 
which opens the possibility of trapping phonons between two breathers.
Numerical results show that 
a perfect reflection occurs at a special value of
$k_{c}$ if the coupling is not too 
strong. 
The value of $k_{c}$ where the perfect reflection occurs changes 
from $\pi / 2$ to $0$ as $\epsilon$ increases.
If $\epsilon$ reaches a critical value of $0.36$, 
the perfect reflection disappears through $k_{c}=0$ as depicted in
Fig.\ref{fig6}.

\section{conclusion}
In conclusion, phonon scattering by discrete breathers
in the DNLS equation is formulated 
by the transfer matrix method and transmissions were obtained in the 
one-channel scattering case.
The nonzero transmission of phonons through
the discrete breather at $k=0$ in the one-channel phonon 
scattering occurs at the localized eigenmode threshold is proven 
analytically 
to be perfect for the symmetric discrete breather.
Numerically, the perfect transmission was shown to occur at $k=0$ 
whenever the symmetric and antisymmetric localized eigenmodes 
were generated from the 
lower edge of the phonon band. 
We obtained explicitly the formula for the phonon transmission
in the lowest order of the coupling strength.
A perfect reflection was shown to exist at $k=\pi/2$ in the small coupling 
limit.
The location of the perfect reflection was traced numerically 
until it disappeared through $k_{c}=0$ at a critical value of 
the coupling strength.

The solution matching method using the transfer matrices 
has been proved to be an efficient tool for investigating scattering properties.
Furthermore, the DNLS system provided a good intermediate example with
two scattering channels 
bridging the gap between the case of a kink in the local equilibria 
with a single scattering channel
and the discrete breather in the classical Hamiltonian system
with an infinity of scattering channels.
The methods used for the breather in the DNLS system can be extended 
to study the one-channel scattering of the nearest neighbor Hamiltonian chain.
The multi-channel scattering problem is important in the context of stability,
e.g.,  a two-channel scattering problem 
where the phonons of the 
second channel are incident from one side of the breather.
Detail study of phonon-induced instability will be the subject of 
on-going investigations.

\section{acknowledgments}
S.L. and S.K. are suppored byt the Korea Science and 
Engineering Foundation
through contract 97-0202-009-2.
J.J.L.T. acknowledge the support of APEC fellowship. 
We are also grateful to S. Flach and S. I. Choi for stimulating discussions.

\begin{figure}
\label{fig1}
\epsfxsize=\columnwidth\epsfbox{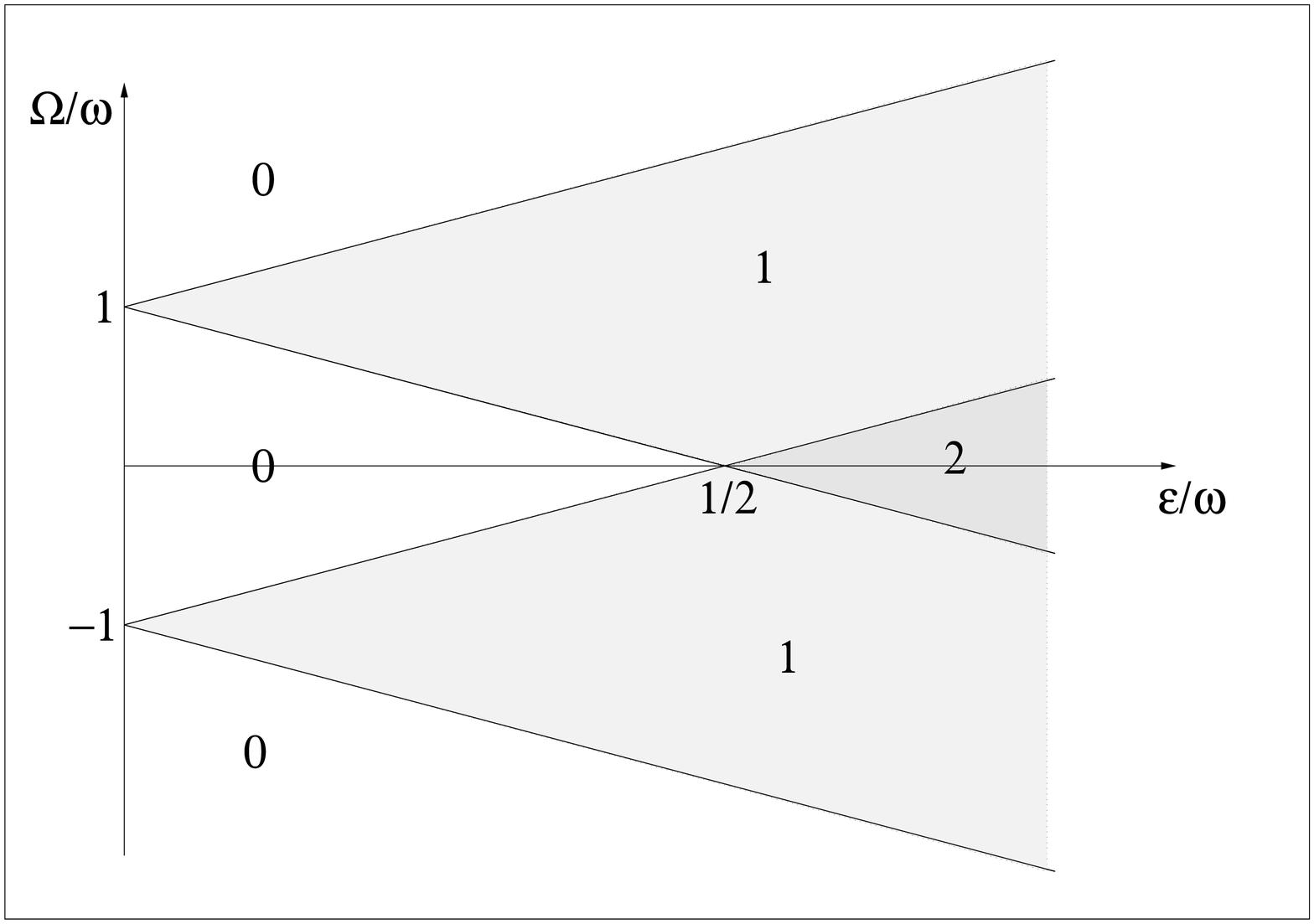} 
\vspace{0.3cm}
\caption{The eigenspectrum of linearized modes of 
the DNLS equation as a function of $\epsilon$. 
The phonon bands are shaded.
The number in each region of the diagram 
denotes the number of the traveling channels. }
\end{figure}


\begin{figure}
\epsfxsize=\columnwidth\epsfbox{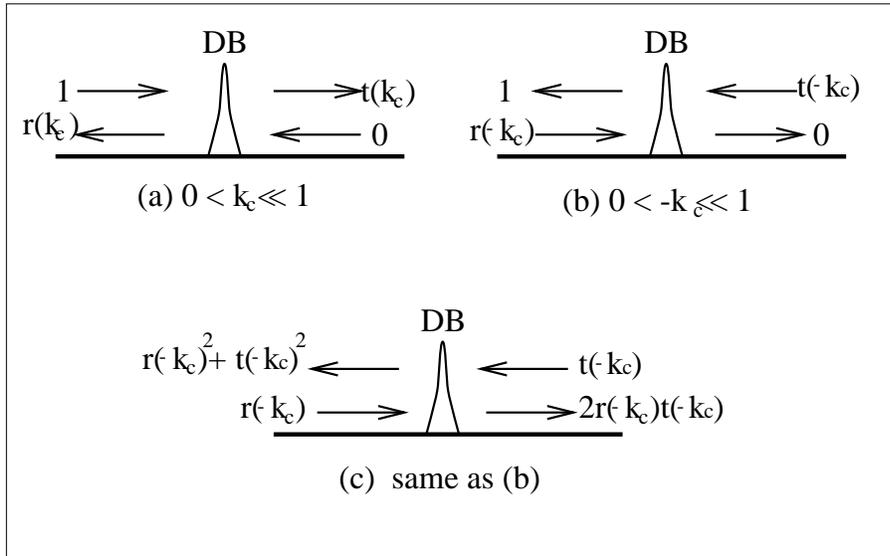} 
\caption{Continuation of the one-channel phonon-breather scattering 
configuration from a small positive value of $k_c$ to a small negative 
one.
}
\label{fig2}
\end{figure}

\begin{figure}
\epsfxsize=\columnwidth\epsfbox{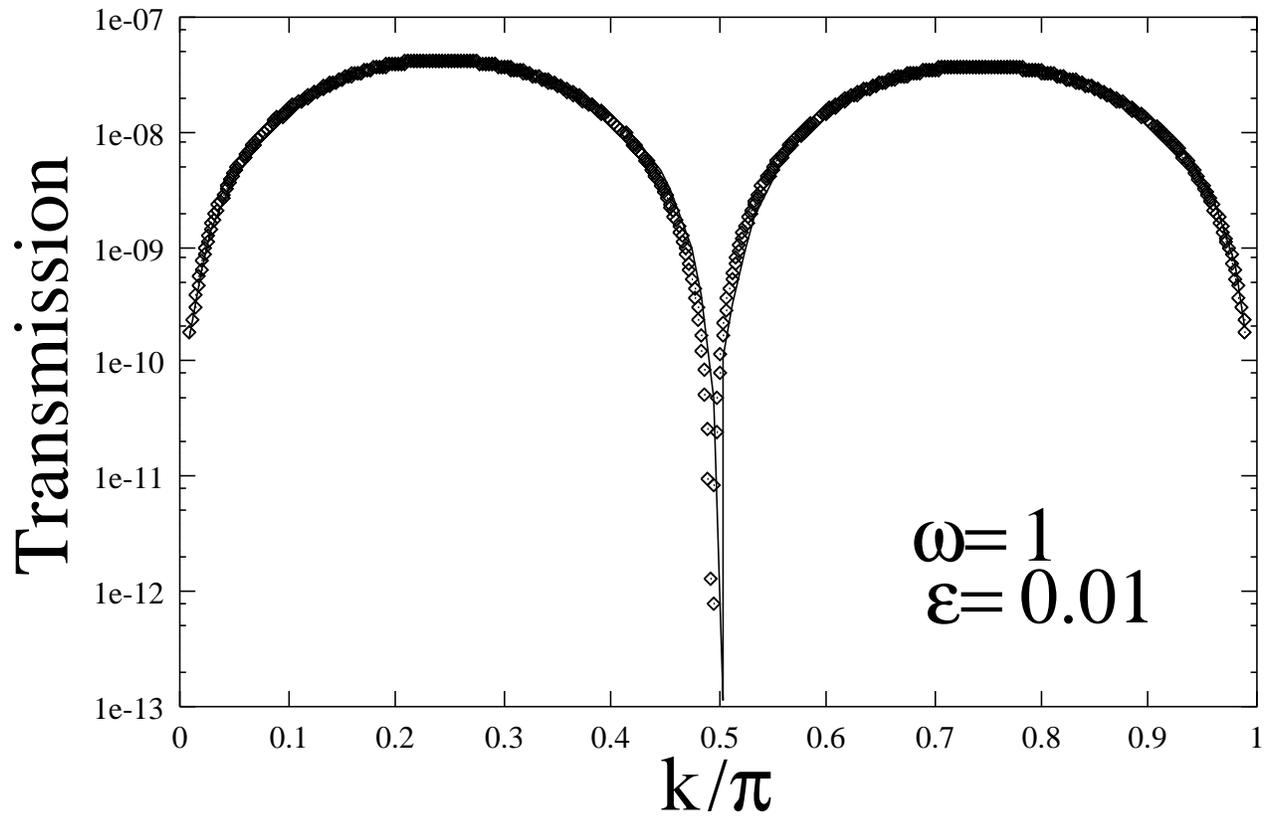} 
\caption{Phonon transmission through the breather in the
DNLS equation in the weak-coupling limit. 
The solid line is the perturbative result, while diamonds denote
the result of numerical calculations using the transfer matrix method.}
\label{fig3}
\end{figure}

\begin{figure}
\epsfxsize=\columnwidth\epsfbox{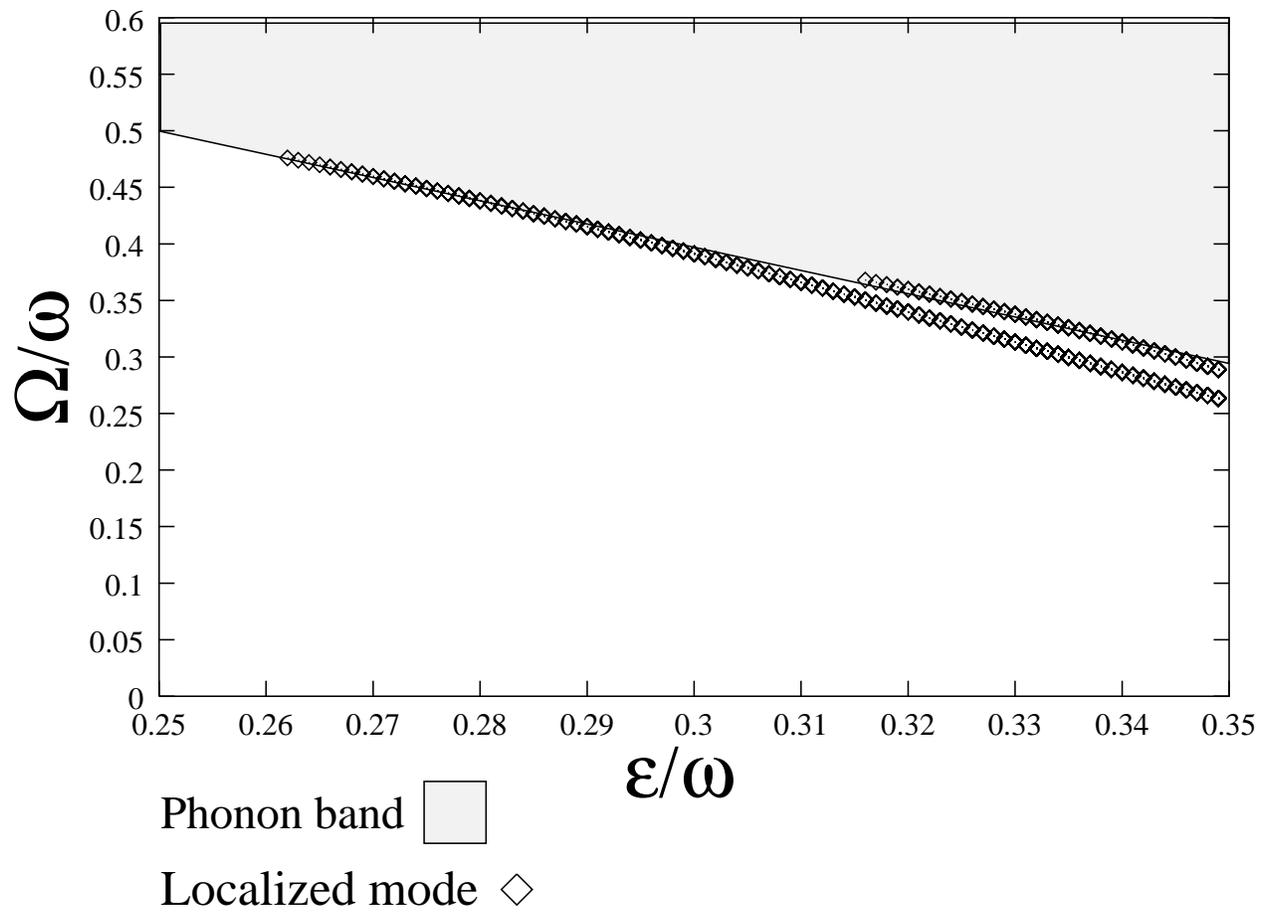} 
\vskip 0.5cm
\caption{
The blowup of Fig.\ref{fig1} with two branches of 
localized eigenmodes coming out of the phonon band.}
\label{fig4}
\end{figure}

\begin{figure}
\epsfxsize=\columnwidth\epsfbox{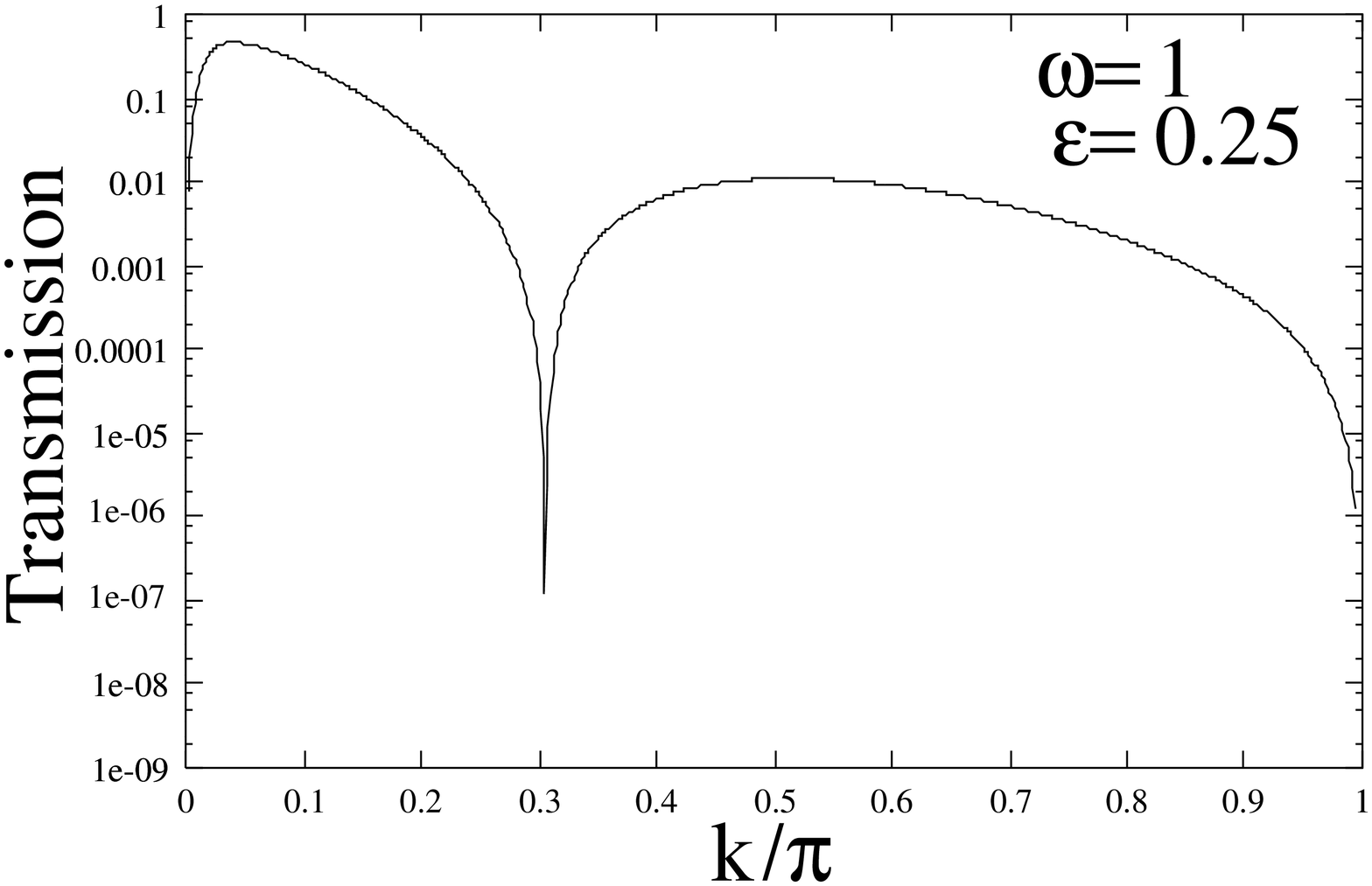} 
\epsfxsize=\columnwidth\epsfbox{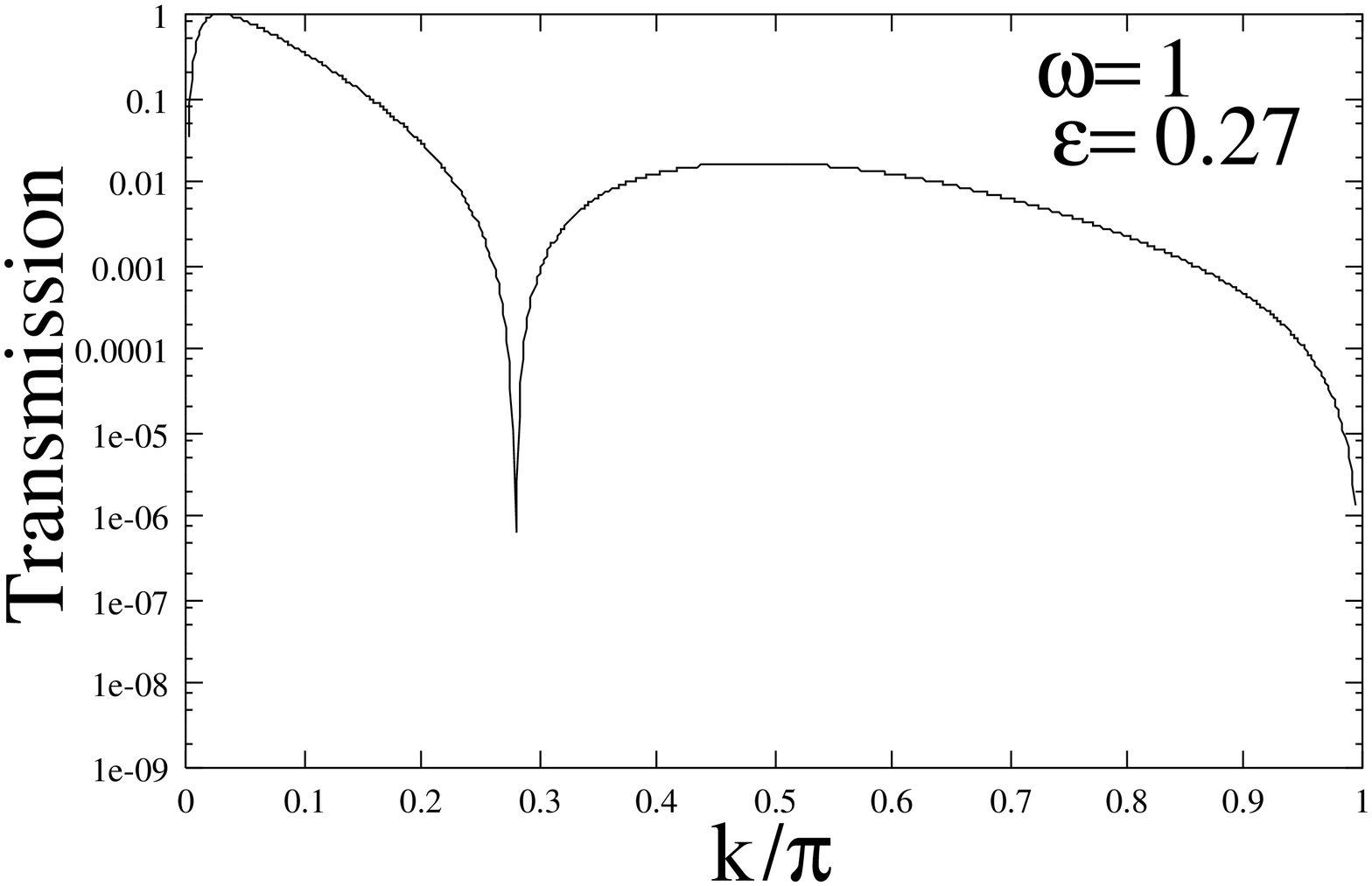} 
\vspace{0.5cm}
\epsfxsize=\columnwidth\epsfbox{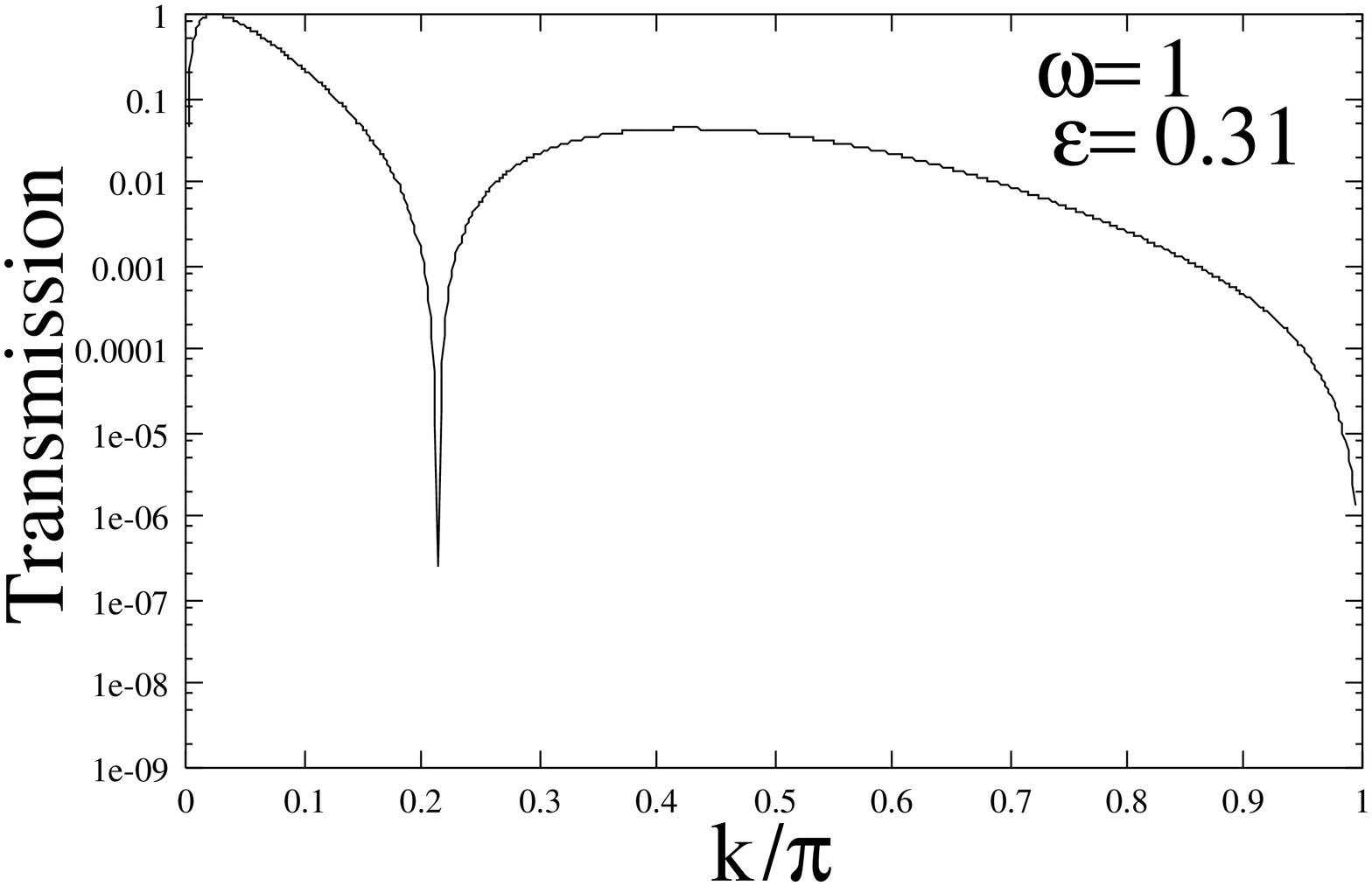} 
\epsfxsize=\columnwidth\epsfbox{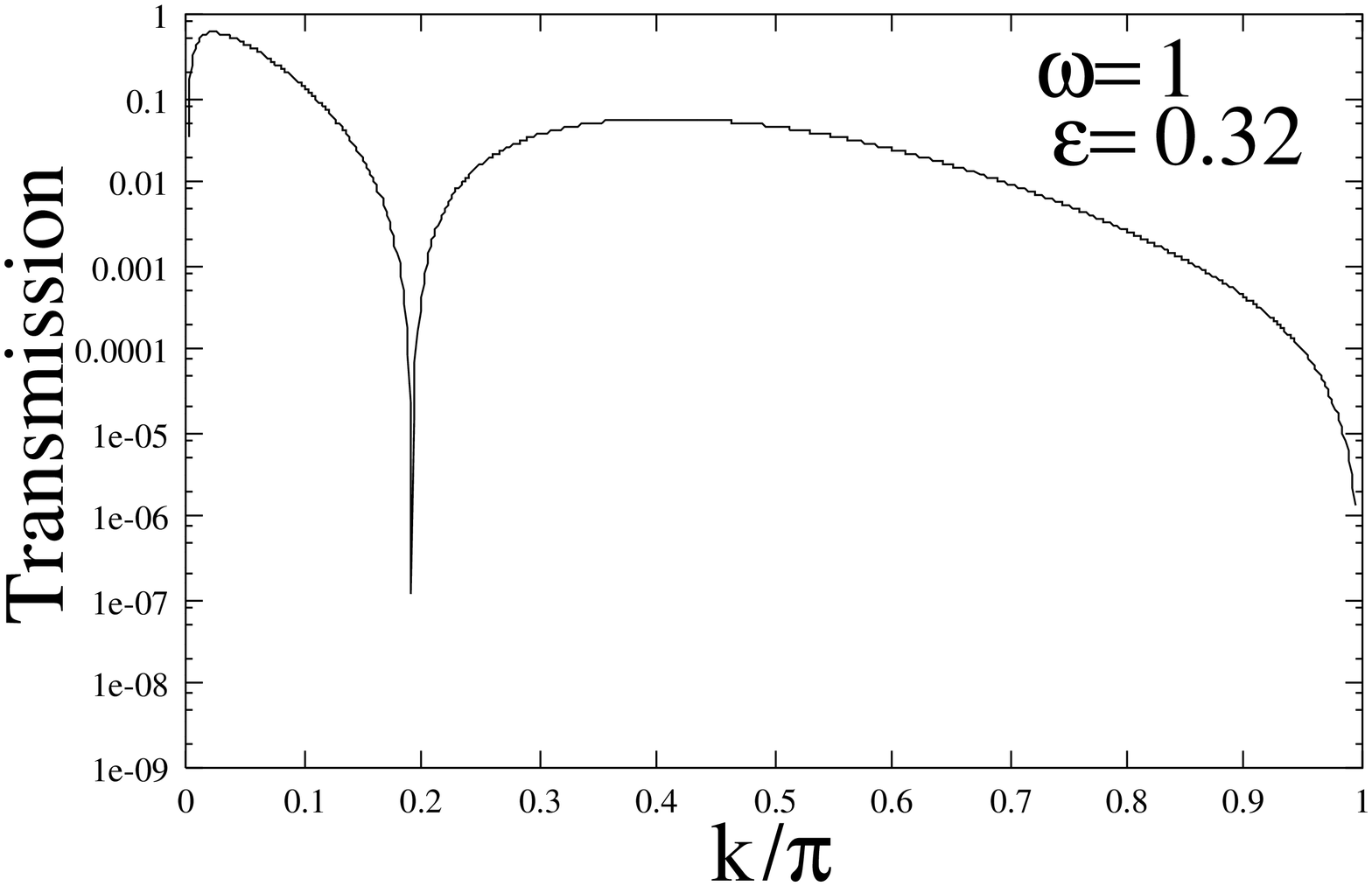}
\vspace{0.5cm}
\caption{One-channel phonon transmissions through discrete breathers in the 
DNLS equation for $\omega=1$. 
(a)-(b) Near $\epsilon=0.262 $
a perfect transmission is generated.
(c)-(d) Near $\epsilon=0.315$,
the perfect transmission is annihilated.}
\label{fig5}
\end{figure}

\begin{figure}
\epsfxsize=\columnwidth\epsfbox{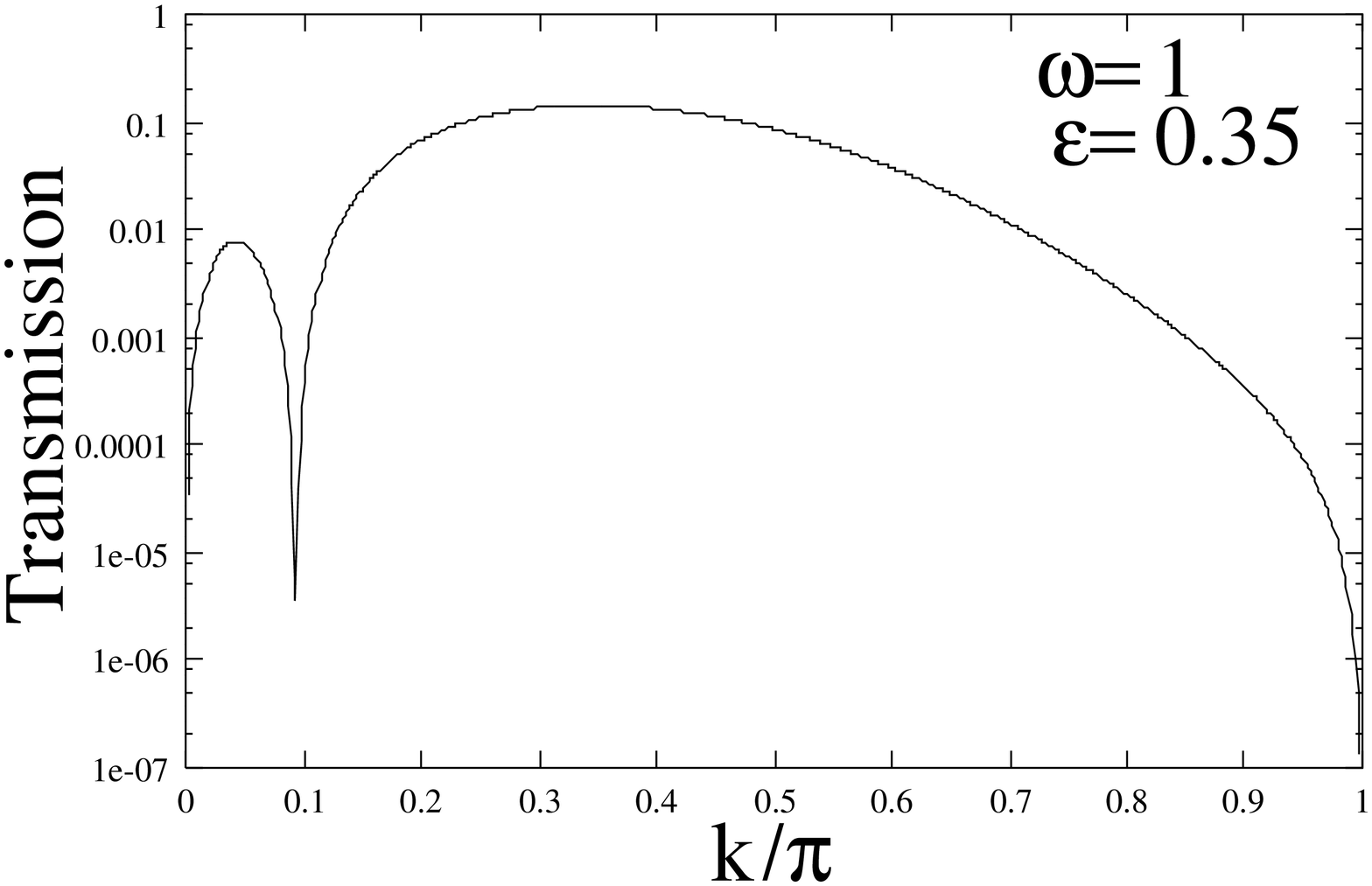} 
\epsfxsize=\columnwidth\epsfbox{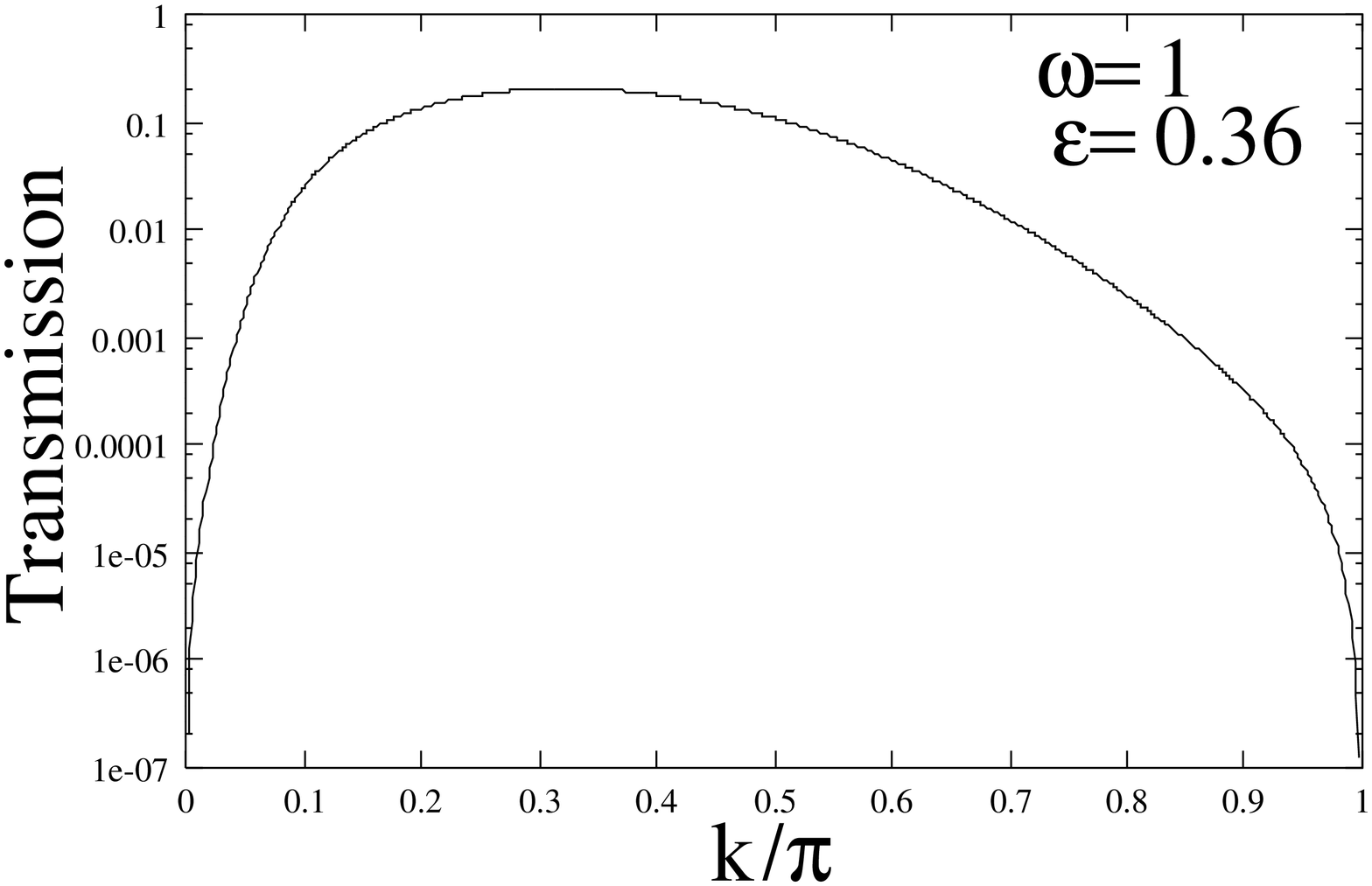} 
\epsfxsize=\columnwidth\epsfbox{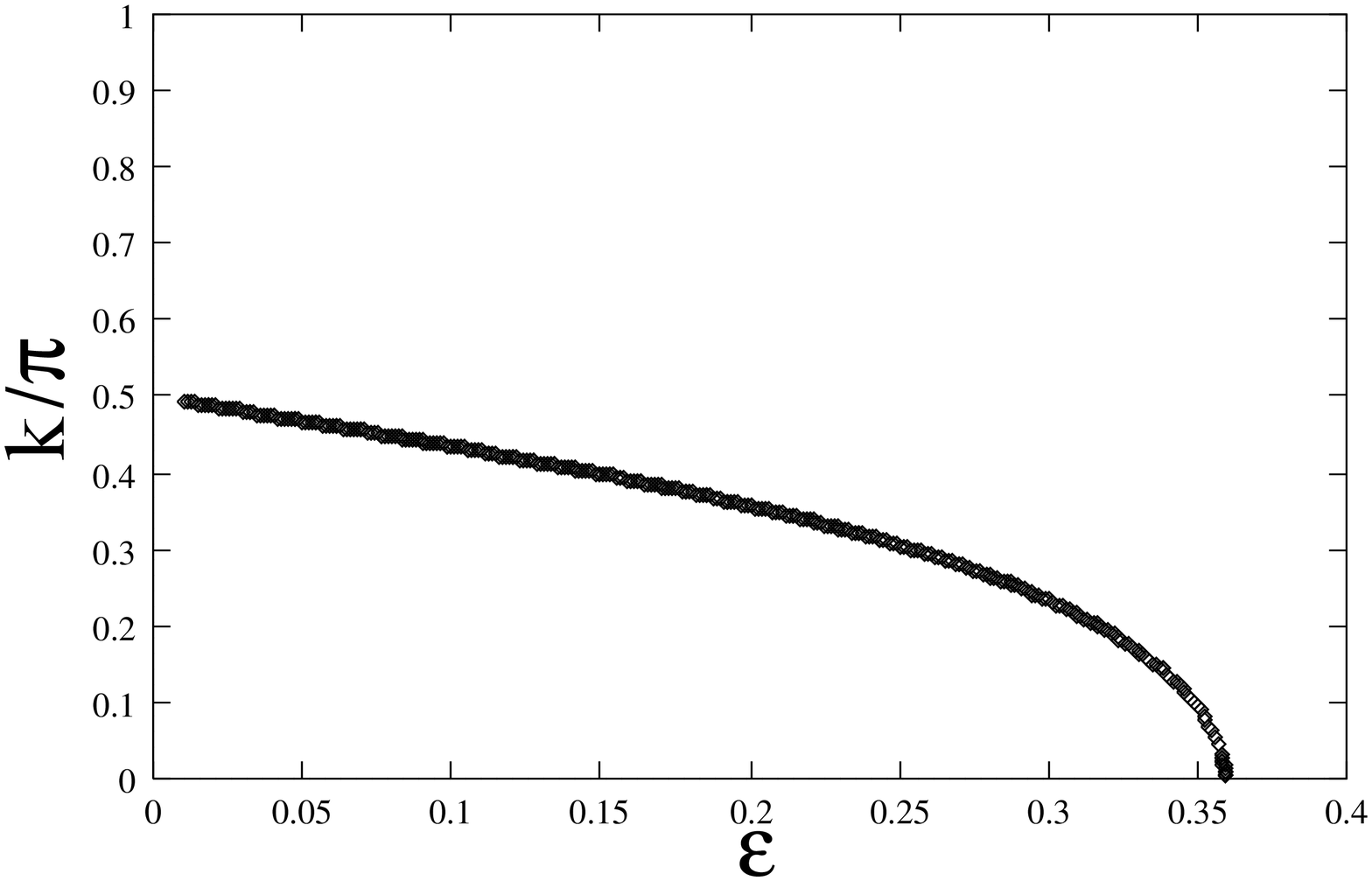} 
\vskip 0.5 cm
\caption{One-channel phonon transmissions through breathers
in the DNLS equation for $\omega=1$. 
Near $\epsilon=0.36$, (a) the perfect reflection occurs at $k \ne 0$ or $\pi$; 
(b) as the coupling strength is increased it is annihilated at $k=0$. 
(c)The wave vector for the perfect reflection as a function of the coupling 
strength $\epsilon$. }
\label{fig6}
\end{figure}
\end{document}